# RAMAN SCATTERING OF WATER AND PHOTOLUMINESCENCE OF POLLUTANTS ARISING FROM SOLID-WATER INTERACTION

Ph. Vallée[1*], J. Lafait[1], M. Ghomi[2], M. Jouanne[3], J.F. Morhange[3].

[1] *Laboratoire d' Optique des Solides (UMR CNRS 7601) – case 80*

[2] *Laboratoire de Physicochimie Biomoléculaire et Cellulaire (UMR CNRS 7033) – case 138,*

[3] *Laboratoire des Milieux Désordonnes et Hétérogènes (UMR CNRS 7603) – case 86*

*Université Pierre et Marie Curie,*

*4, place Jussieu, 75252 Paris Cedex 05, France*

*Corresponding author: phvallee@los.jussieu.fr

**Abstract**

Systematic Raman experiments performed on water and water-ethanol samples, stored in different containers (fused silica, polypropylene, soda-lime glass type III) for several hours, have shown that the luminescence contribution to the Raman signal fluctuations is directly related to the container composition. Intensity fluctuations as large as 98%, have been observed in the spectral regions corresponding to the both water intramolecular and intermolecular vibrations, despite the fact that the wavenumbers of the modes remained unchanged. We undoubtedly attribute these fluctuations to a luminescence phenomenon on the basis of : i) the absence of such effect in the anti-Stokes domain, ii) its dependence on the excitation laser wavelength, iii) other relevant photoluminescence experiments. This luminescence is attributed to pollutants at ultra-low concentration coming from the different containers.

# 1. Introduction

Our aim is to give an additional contribution to the question of "intrinsic luminescence" of water already addressed by Lobyshev et al. [1, 2]. Independently of a possible interpretation in terms of intrinsic luminescence, few authors [3, 4] have published results pointing out the luminescence effects in water. Most of these effects were finally attributed to the role of admixtures. Moreover, it is well known that luminescence and Raman spectroscopy performed on liquids, gases or solids, are highly sensitive to traces of pollutants, even at ultra-low concentrations. These techniques are already used for the detection of pollutants in water.

The objective of the present work is to bring a contribution in analysing tiny and subtle effects of electromagnetic waves on water. Raman scattering can be considered as a relevant technique for studying these effects which are supposed to act on the intensity of the scattering intensity and not on the wavenumber of the characteristic bands of water molecule vibrations. It should be recalled that those working on aqueous solutions often calibrate their experiment on the Raman bands of water. We therefore focused our attention on the reproducibility of the intensity of water Raman bands. At this step, we observed large variations of the Raman response that we had to correlate with some basic phenomena and our experimental process.

The first aim of the present report is to demonstrate the occurrence of luminescence phenomena due to pollutants coming from the tubes containing the water samples used generally for experiments. The second aim is to question the interpretation of Lobyshev's experiments in terms of intrinsic luminescence of water.

After the presentation of the characterisation techniques used for our experiments (section 2), we describe and discuss in section 3 our results obtained from pure water and on water-ethanol samples.

## 2. Experimental

In order to achieve our first aim (see above, Introduction), we used two kinds of liquid samples: pure water and hydro-alcoholic solution. Adding ethanol to water increases the apolarity of the liquid and therefore enhances the solubilization of organic components possibly present in the wall of the container. Raman experiments were completed with photoluminescence experiments at a different excitation wavelength and the conductivity of the samples has been measured at different steps of the experiment. All data were recorded at room temperature (20-25°C)

*Liquids for samples*

- pure water : SEROMED produced by Biochrom KG, apyrogen, conductivity between 0,06 and 0,08 $\mu S\ cm^{-1}$ directly after the production and stored in bottles of borosilicate glass type I, passivisated in order to reduce the migration of its components in the content.
- hydro-alcoholic solution: 50% ethanol Chromanorm Merck and 50% SEROMED water

*Containers*

Liquids were stored during twelve hours in three different kinds of containers :

- Soda-lime glass type III disposable operculated tubes (Prolabo)
- Polypropylene disposable operculated tubes (Polylabo)
- Quartz glass Suprasil cells (Heraeus)

For the optical experiments (Raman scattering and photoluminescence), liquids were transferred in Quartz Suprasil cells of 3,5 ml inner volume (Hellma)

*Raman scattering experiment*

Raman scattering was excited with an Argon laser (Stabilite, Spectra Physics, $\lambda$=514.5nm, power level of 100 mW focused with a 100µm diameter beam in the middle of the cell) and collected, in 90° configuration, on a Jobin-Yvon U1000 spectrometer double monochromator equipped with a photomultiplier counter cooled by a Peltier cell . Stokes and anti-Stokes spectra were recorded between –500 and +500 $cm^{-1}$ in order to distinguish Raman lines from luminescence. The total range of measurement was –500 to +4000 $cm^{-1}$. At low frequency, all spectra were reduced by the Bose-Einstein factor [5, 6]. The power incident on the sample was measured before and after each scan.

*Photoluminescence experiment*

The photonic excitation was realized at 488 nm with an Argon laser (Coherent, Innova 300, 100 mW). Light emitted from the sample was collected, in 90° configuration, by an optical fiber located at 10 mm of the cell and analyzed by a Jobin-Yvon Spectrometer HR 460 equipped with a multichannel CCD detector (Spectraview-2D, 2000 pixels). The resolution is 4nm/point with a 300 µm slit.

*Conductivity measurements*

Conductivity measurements were realized using a Multi-meter Consort C835, with an electrode of cell constant K=0,1 $cm^{-1}$, especially adapted for low ionic concentration solutions, with automatic temperature correction (Pt1000).

## 3. Results and discussion

We recall that with the aim of demonstrating the influence of incidental pollutants on water samples, we decided to use pure water samples and 50% water – 50% ethanol sample;. the latter samples showing a higher capability for organic compound dissolution.

*a) pure water samples*

We first present the results obtained with samples of pure SEROMED water (see above, section 2) stored in three different kinds of tubes : quartz glass (QW), polypropylene (PW) and soda-lime glass type III (GW). Three kinds of experimental results are compared : Raman scattering, photoluminescence and conductivity measurements.

Figures 1, 2 and 3 show the Raman spectra of these three samples excited at 514.5 nm, whereas Table 1 gives the increase (in relative value) of the Raman intensity of PW and GW samples compared to that of QW sample considered as a reference, at discrete frequencies, characteristic of the intramolecular and intermolecular vibrations of the water molecules.

The Raman spectrum of GW samples clearly shows a large intensity excess over those from QW and PW spectra, especially in the middle (Figure 2) and high (Figure 1) frequency ranges of the Stokes domain. This excess can be as large as 98% at *ca.* 2110 cm$^{-1}$ (Table 1). Then this excess decreases slowly both in the middle and high frequency ranges (respectively 60% at 1640 cm$^{-1}$ and 23 % at 3240 cm$^{-1}$). At lower frequencies and in the anti-Stokes domain (Figure 3), this difference never exceeds 16%. Considering this small excess value at low frequency region and the fact that it is not observed in a symmetric manner in the Stokes and anti-Stokes domains (see Table 1: 14% at –470 cm$^{-1}$ and 28% at +470 cm$^{-1}$) we assign this observation to the photoluminescence effect.

With this hypothesis, we performed photoluminescence experiments on the two most characteristic samples (GW and QW). We excited this luminescence at 488 nm (instead of 514.5) in order to demonstrate the dependence of this effect with the energy of excitation. And effectively, the luminescence response of GW sample excited at 488 nm is extremely high around 2000 cm$^{-1}$; it reaches 4.5 times the QW response at 2110 cm$^{-1}$ (544nm), instead of 98% at 2110 cm$^{-1}$ (577nm) when excited at 514.5 nm (see Figure 4 and Table 1). This strong photoluminescence effect, absent (or very weak) in the QW sample, can only be attributed to the luminescence of pollutants. In order to verify this assumption, we measured the conductivity of our samples (see table 2). The conductivity of GW samples is effectively four times larger than that of QW and PW samples, whereas their pH is roughly the same: 6.8.

Therefore, one can conclude that pollutants, in very tiny concentration, coming probably from the walls of the tubes (all manipulated in the same conditions) are responsible for the photoluminescence signal disturbing the intrinsic Raman scattering spectrum of water. This signal is evidently frequency dependent and is particularly high when excited at 488 nm. The main contribution of this photoluminescence is in the intramolecular vibration domain of the Raman spectrum of water. Its maximum occurs at *ca.* 2110 cm$^{-1}$, frequency associated to the additive combination of the of the water molecule bending mode (*ca.* 1640 cm$^{-1}$) and of an intermolecular mode (*ca.* 470 cm$^{-1}$), [6].

*b) Water-ethanol samples*

The same experiments were performed on 50% water – 50% ethanol samples in order to enhance the apolarity of the liquid and point out the possible differences of behaviour in the different kinds of containers.

Figure 5 presents the Raman spectra of the three samples : 50% water – 50% ethanol in : quartz glass (QWE), polypropylene (PWE) and soda-lime glass type III (GWE). The spectrum of pure SEROMED water in quartz glass (QW) has been added for comparison. As compared to Figure 1 (and to curve QW), a lot of new bands show up in curves QWE, PWE and GWE, in the frequency range 300-3000 cm$^{-1}$, characteristic of the vibrations of the ethanol molecule. The background envelope of the ethanol vibration modes clearly departs from the base response of pure water (curve QW) and can be attributed to photoluminescence of specific impurities present in ethanol and of pollutants coming from the tubes and specifically solubilized by ethanol. It is obvious, by comparing the responses of samples QWE and QW, of respectively water-ethanol and pure water in tubes of the same composition (quartz glass). This extra luminescence contribution to the Raman spectrum is important when excited at 514.5 nm and extends from 400 to 2700 cm$^{-1}$, with a maximum around 1360 cm$^{-1}$ (553 nm). An other evidence of the luminescent character of this contribution can be given by varying the laser power of the Raman excitation. By decreasing power from 100 to 50 mW, one can observe total vanishing of this contribution (see Figure 7 in the case of soda-lime glass container), while the Raman modes remain unchanged.

When excited at 488 nm in a photoluminescence experiment (see figure 6), the maximum of the photoluminescence spectrum now occurs around 2100-2500 cm$^{-1}$ (543-556 nm), like in the previous experiment with pure SEROMED water (see figure 4), while the extra contribution previously pointed out with a maximum at *ca.* 1360 cm$^{-1}$ is considerably weakened, but still present (see Figure 6 and Table 1). The frequency dependence of these features supports their interpretation in terms of photoluminescence effects.

Moreover, conductivity measurements performed on QWE, PWE and GWE samples (see Table 2), qualitatively confirm these interpretations. The basic conductivity of QWE and PWE samples are equal and roughly two times those corresponding to QW and PW samples. It is due to a sum of the intrinsic conductivity of the water-ethanol solution, and to the ionic impurity conductivity of ethanol. These impurities contribute to the main broad photoluminescence band pointed out above. On the other hand, the GWE conductivity is smaller than the GW one (4.6 compared to 6.7 $\mu Scm^{-1}$). The GWE conductivity results from two contributions: the previously mentioned ionic impurity conductivity of ethanol plus the pollutant conductivity coming from dissolving of soda-lime glass components in the water-ethanol solution. The second contribution can be weaker in GWE samples than in GW ones due to weaker dissolution of soda-lime glass components in the water-ethanol solution than in pure water. This explanation is supported by the weaker photoluminescence ratio of these pollutants observed at 2110 $cm^{-1}$ (0.44 instead of 4.48, see Table 2).

*c) Complementary remarks*

- Raman scattering and photoluminescence spectra presented above were reproduced on a significant number of samples prepared in the same conditions.
- Conductivity values presented in Table 2 result from averaging on conductivity measurements performed on these different samples.
- Moreover, photoluminescence spectra of samples prepared with demineralized water of lower conductivity than SEROMED water (0.5 $\mu Scm^{-1}$ instead of 1.6 $\mu Scm^{-1}$), were identical to those prepared with SEROMED water (Figure 4) for both quartz glass and soda-lime glass containers. This result confirms, (i) that it is really the luminescence of the components of the container which is observed, (ii) that dissolving of these components is the same by demineralized water as by SEROMED water.

## 4. Conclusions

By the joint use of Raman scattering, photoluminescence and conductivity experiments, we have shown that the study of the optical properties of high purity water samples can be strongly perturbed by the photoluminescence of traces of pollutants coming from the container. Classical soda-lime glass type III tubes are clearly the worse containers. Pure water (SEROMED or demineralized) stored in polypropylene tubes exhibits very weak photoluminescence, slightly higher than that in quartz glass tubes. On the other hand, alcoholic solutions stored in polypropylene tubes exhibit a much higher photoluminescence than in quartz glass tubes. Moreover polypropylene due to its composition may release organic components. In conclusion, quartz glass is clearly the most suitable container material for experiments involving high purity water as it minimizes content-container interactions.

Other characterisations like Inductively Coupled Plasma Mass Spectrometry (ICP-MS) are envisaged in order to identify the pollutants which induce the photoluminescence observed in our experiments. We also plan to perform photoluminescence excitation spectra for a better characterization of the pollutant luminescence.

At this step of our experiments, in the absence of the complementary characterizations mentioned above, it seems difficult to definitely conclude on a possible intrinsic luminescence of water, already experimentally demonstrated by Lobyshev et al [1, 2]. Nevertheless, for the moment, we have demonstrated the difficulty to obtain water samples totally free of any pollutants capable to produce tiny noisy photoluminescence effects.

**Acknowledgements**

This work is part of a PhD thesis prepared by Ph. Vallée, supported by a grant of the Fondation Odier which is gratefully acknowledged by the person concerned. The scientific committee following this work is also warmly thanked : P. Mentré, M.O. Monod, R. Strasser, Y. Thomas. B. Démarets is thanked for his help in mechanical realisations.

**References**

[1]  V.I. Lobyshev, R.E. Shikhlinskaya and B.D. Ryzhikov, J. Mol. Liq., 82 (1999) 73.

[2]  V.I. Lobyshev, R.E. Shikhlinskaya, B.D. Ryzhikov and T.N. Mazurova, 39 (1994) Biofizika, 565.

[3]  S.V. Patsaeva, EARSEL Advances in Remote Sensing, 3 (1995) 66.

[4]  P. Vigny, PhD thesis, Université Pierre et Marie Curie, Paris 6, (1974).

[5]  O. Faurskov Nielsen, Annual Reports of the Royal Society of Chemistry, C 93 (1993) 57.

[6]  S.E. May Colaianni and O. Faurskov Nielsen, J. Mol. Struct., 347 (1995) 267.

**Tables**

Table 1:

| Samples | Exp. | ν (cm⁻¹) | | | | | | | |
|---|---|---|---|---|---|---|---|---|---|
| | | -470 | -165 | 165 | 470 | 1360 | 1640 | 2110 | 3240 |
| **(PW-QW)/ QW** | R | 0.04 | 0.08 | 0.1 | 0.09 | - | 0.15 | 0.13 | 0.19 |
| **(GW-QW)/ QW** | R | 0.14 | 0.14 | 0.16 | 0.28 | - | 0.6 | 0.98 | 0.23 |
| | PL | - | - | - | - | - | 1.64 | 4.48 | 0.07 |
| **(PWE-QWE)/QWE** | R | * | * | * | 0.25 | 1.00 | 1.08 | 1.03 | 0.02 |
| **(GWE-QWE)/QWE** | R | * | * | * | 0.48 | 2.86 | 3.42 | 3.33 | 0.05 |
| | PL | - | - | - | - | 0.05 | 0.17 | 0.44 | 0.02 |

Table 1 : Relative excess of the Raman (R) (excited at 514.5 nm) or photoluminescence (PL) (excited at 488 nm) response of different samples of SEROMED water (W) or water-ethanol (WE) contained in different tubes : quartz glass (Q), polypropylene (P), soda-lime glass (G), at discrete wavenumbers ν (cm⁻¹), over the response of the liquid in quartz tube taken as a reference . * means: relative difference lower than the experimental accuracy.

Table 2

| Samples | Conductivity (μS/cm) |
|---------|----------------------|
| QW      | 1.6                  |
| PW      | 1.5                  |
| GW      | 6.7                  |
| QWE     | 3.0                  |
| PWE     | 3.0                  |
| GWE     | 4.6                  |

Table 2 : Conductivity of different samples of SEROMED water (W) or water-ethanol (WE) contained in different tubes : quartz glass (Q), polypropylene (P), soda-lime glass (G)

**Figure captions**

Figure 1 : Full Raman (excitation 514.5 nm) scattering spectra of SEROMED water samples in quartz glass (QW), polypropylene (PW) and soda-lime glass type III (GW) tubes.

Figure 2 : Expansion of Raman spectra of Figure 1 in the middle frequency domain (1200 to 2500 cm$^{-1}$).

Figure 3: Expansion of Raman spectra of Figure 1 in the low frequency domain (-500 to 1200 cm$^{-1}$).

Figure 4 : Photoluminescence (excitation 488 nm) spectra of SEROMED water samples in quartz glass (QW) and soda-lime glass type III (GW) tubes. Upper abscissa scale refers to wave numbers reduced by excitation wave length.

Figure 5 : Full Raman (excitation 514.5 nm) scattering spectra of : 50% SEROMED water – 50% ethanol solutions in quartz glass (QWE), polypropylene (PWE) and soda-lime glass type III (GWE) tubes and also of SEROMED water sample in quartz glass (QW) tube.

Figure 6 : Photoluminescence (excitation 488 nm) spectra of : 50% SEROMED water – 50% ethanol solutions in quartz glass (QWE) and soda-lime glass type III (GWE) tubes and also of SEROMED water sample in quartz glass (QW) tube and of an empty quartz tube (Q). Upper abscissa scale refers to wave numbers reduced by excitation wave length.

Figure 7 : Raman (excitation 514.5 nm) scattering spectra of : 50% SEROMED water – 50% ethanol solutions in soda-lime glass type III (GWE) tube for two different excitation powers : 50 mW and 100 mW.

**Figures**

Figure 1

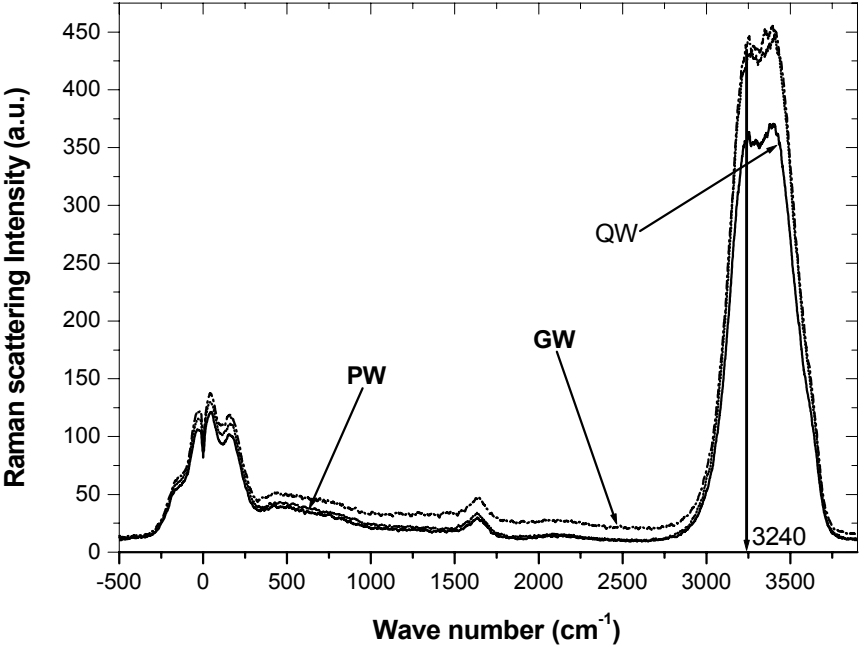

Figure 2

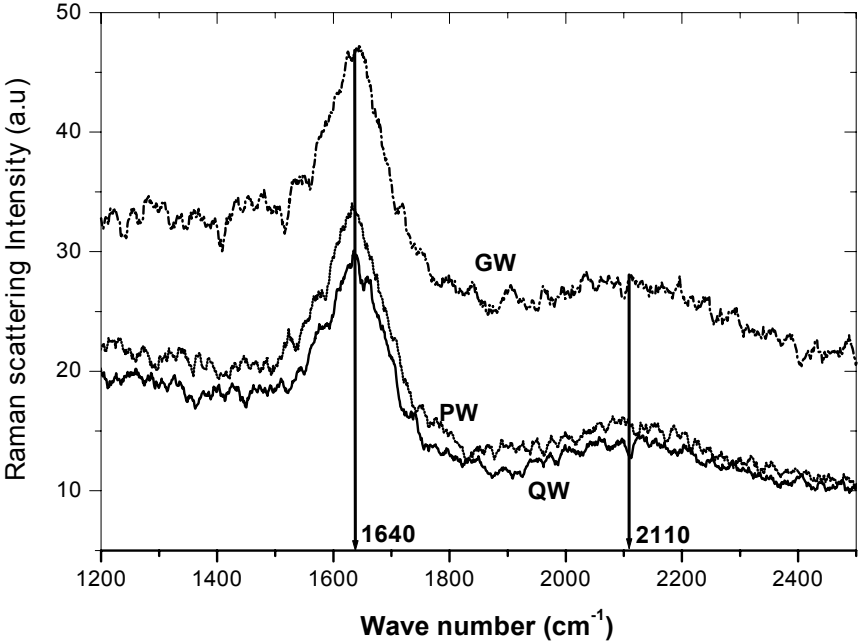

Figure 3

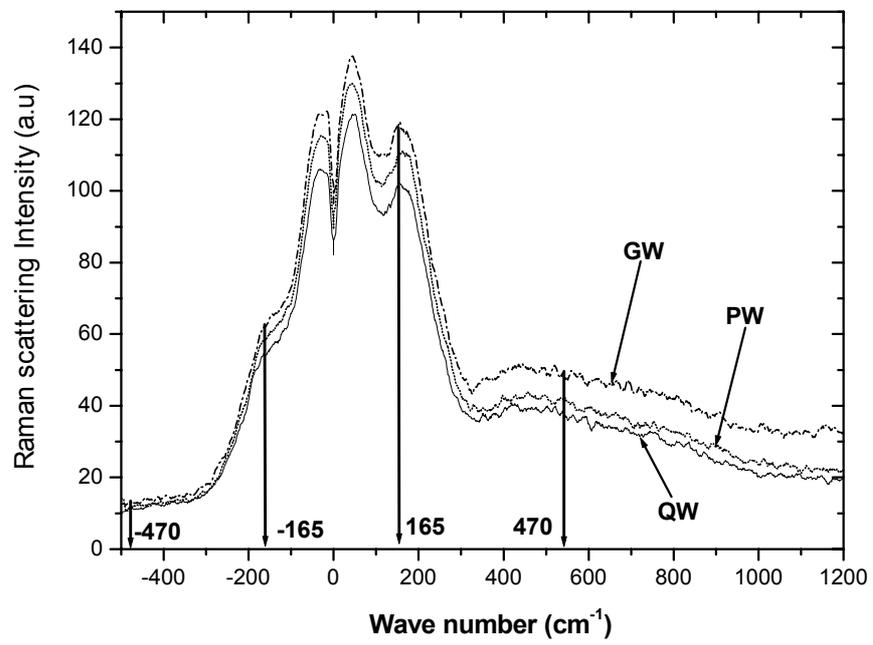

Figure 4

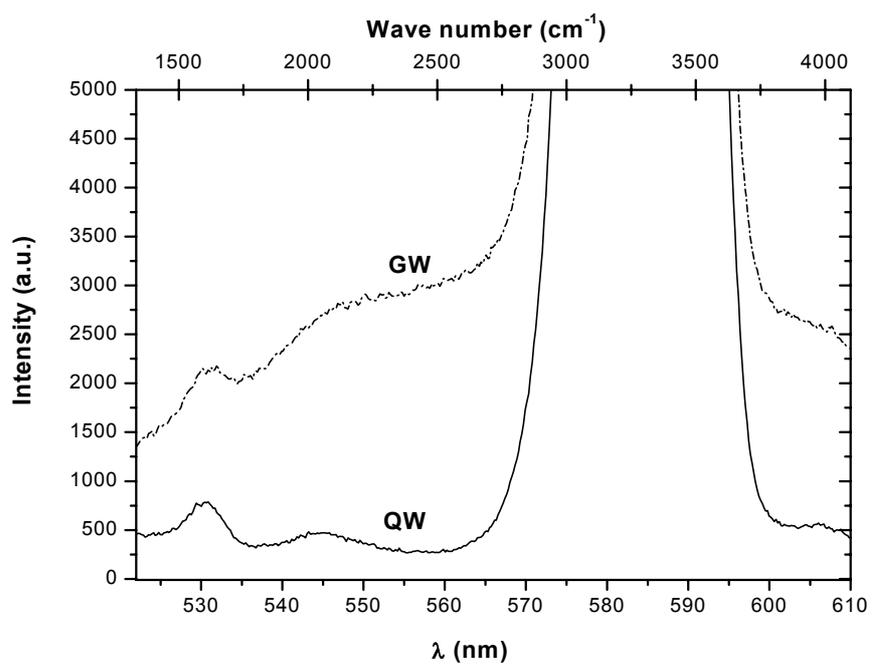

Figure 5

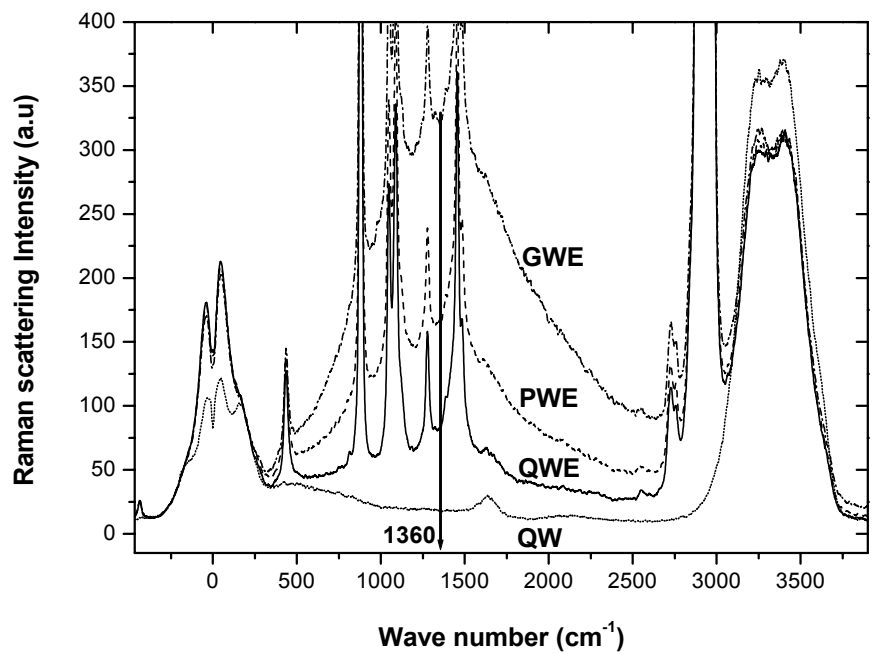

Figure 6

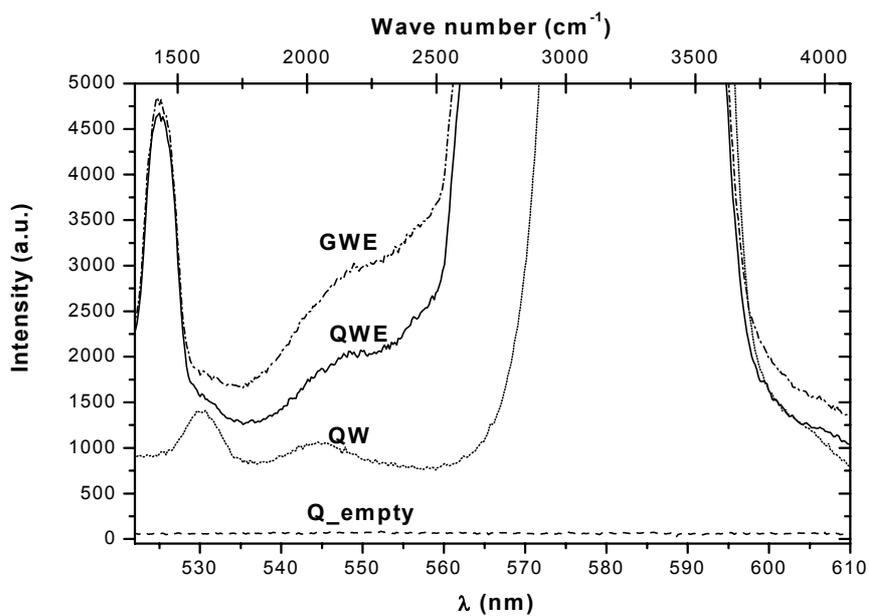

Figure 7

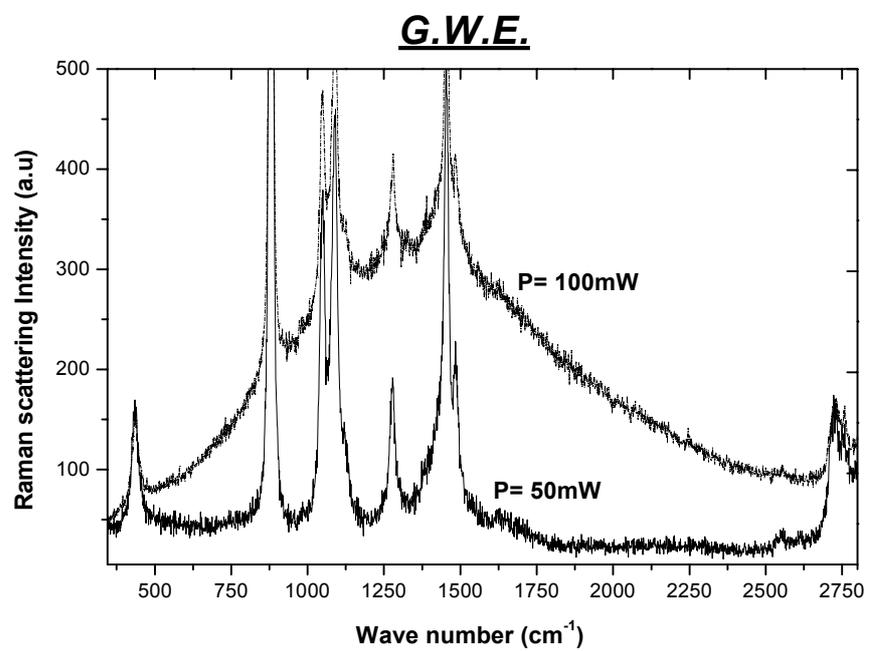